\documentclass[aps,pra,showpacs,notitlepage, 10pt]{revtex4-1}
\usepackage{color}
\usepackage{graphicx}
\usepackage{natbib}
\usepackage{hyperref}
\usepackage{braket}
\usepackage{bm}
\usepackage{amsmath}

\newcommand{\eqaref}[1]{Eq.~\eqref{#1}}

\begin{document}
\title{Interaction of electron vortices and optical vortices with matter and processes of orbital angular momentum exchange}
\author{S. M. Lloyd}
\author{M. Babiker}
\author{J. Yuan}
\affiliation{Department of Physics, University of York, Heslington, York, YO10 5DD, UK}
\pacs{41.85.-p, 42.50.Tx}

\begin{abstract}
The quantum processes involved in the interaction of matter with, separately, an electron vortex (EV) and an optical vortex (OV) are described, with matter modelled in terms of a neutral two-particle  atomic system, allowing for both the internal (electronic-type) motion and the gross (center of mass-type) motion of matter to be taken into account. The coupling of the atomic system to the EV is dominated by Coulomb forces, while that of the OV is taken in the $\mathbf{p}\cdot\mathbf{A}$ canonical form which couples $\mathbf{A}$, the transverse vector potential of the optical vortex, to the linear momenta of the two-particle system.  An analysis of the dipole active transition matrix element is carried out in each case. The electron vortex is found to be capable of exchanging its orbital angular momentum (OAM) with both the electronic and the center of mass motions of the atomic system in an electric dipole transition.  In contrast, for electric dipole transitions the optical vortex is found to be capable of exchanging OAM only with the center of mass.  The predictions are discussed with reference to recent experiments on electron energy loss spectroscopy (EELS) using EVs traversing magnetised iron thin film samples and those involving OVs interacting with chiral molecules. 
\end{abstract}

\maketitle

\section{Introduction}
Vortices are now well established phenomena in a number of physical contexts, most notably in condensed matter physics, where they are known to play an important role in the dynamics of superfluid flow \cite{Pines1994} and, more recently, in dilute gas Bose-Einstein condensates \cite{Fetter2001}; in optics as optical vortices in the form of Laguerre-Gaussian and Bessel light beams \cite{Allen1999b, Allen2003, Andrews2008, Grier2003} and, very recently, in electron microscopy as electron vortices \cite{Uchida2010, Verbeeck2010, McMorran2011}. Broadly, a vortex field has a propagating wavefront endowed with a screw dislocation.  The distinctive feature throughout is the presence of a phase factor $\exp{(il\phi)}$ where $\phi$ is the azimuthal angle about the beam axis, and $l$, an integer taking positive and negative values, is the winding number, such that the field posesses orbital angular momentum (OAM) of $l\hbar$.  Over the last two decades or so, much work has been carried out on optical vortex (OV) beams, so much so that optical vortex physics has now been established as  a new branch of modern optics.  The area began with the seminal work by Nye and Berry \cite{Nye1974} who put forward the suggestion of beams endowed with the vortex property, and the work  by Allen \textit{et al}.~\cite{Allen1992}, first reported in 1992, which set the scene for the subsequent research into optical vortices and their interaction with matter.  Optical vortex physics is rich in both fundamental considerations and applications, and has led to key applications in the manipulation of matter, both in the bulk and at the level of basic constituents.  This impact is envisaged to continue in inter-disciplinary areas, including quantum information processing.

The phenomenon of electron vortices (EVs) is a very recent addition to the growing catalogue of vortex physics, essentially arising from the concept of OVs. The suggestion for their existence was first made by Bliokh \textit{et al}.~\cite{Bliokh2007}. Like an OV, an EV also carries the key property of quantized OAM of $l\hbar$ per electron.  Following Bliokh \textit{et al}.'s suggestion, experimental work led to the creation of EVs,  first by Uchida and Tonomora \cite{Uchida2010} using a stepped spiral phase plate, followed by Verbeeck \textit{et al}.~\cite{Verbeeck2010} who used a binary holographic grating with a Y-like point defect, and thirdly by McMorran \textit{et al}.~\cite{McMorran2011} who also used the holographic plate technique but managed to generate EV beams with winding numbers as high as $l=100$. It is now clear that EVs can be readily generated inside an electron microscope.  More recent advances include the creation of EV beams of cross sections in the atomic scale \cite{Verbeeck2011} and the realisation that they can also be generated using spiral phase plates \cite{Verbeeck2012}.  EV beams are predicted to lead to important new advances in the physics  and potential inter-disciplinary applications of electron beams, revolutionizing electron microscopy and spectroscopy with atomic scale resolution in the imaging of materials, including those with low absorption contrast such as biological specimens \cite{Jesacher2005}. 

Since OAM is a well defined quantized property both of OVs and EVs, it is natural to expect it be exchanged when either type of vortex beam interacts with matter in the form of atoms, molecules or solids. Both optical spectroscopy using OV beams and electron energy loss spectroscopy  (EELS) using EV beams are expected to involve an exchange of quantized OAM with matter, especially in the normally dominant electric dipole transitions.  The exchange must involve both the `electronic-type' motion and the `center of mass-type' motion of matter.   The purpose of this paper is to systematize the theory appropriate for the processes of OAM exchange in the interactions of OVs and EVs, separately,  with a model system of matter in the form of a neutral two-particle system. The main aim is to find out whether, and if so, in what manner, do EVs differ from OVs in the processes of exchange of OAM. This is done by extracting information residing in the respective transition matrix elements leading to the selection rules associated with OAM transfer.\newline
\indent This paper is organised as follows.  In section \ref{OVEVMF} we outline the basic formalisms for the OVs and EVs in the forms of free Bessel modes, define the model atomic system and emphasise the necessity of separating the matter motion into internal (electronic-type) and gross (center of mass-type) motions.  We write down the total Lagrangian of the system relative to the laboratory frame and follow canonical steps to determine the conjugate momenta, leading to the total Hamiltonian including the respective interactions, which we then restrict to the dipole approximation.  Finally we determine the unperturbed quantum states participating in a typical transition. In section \ref{OVCase} we consider OVs in interaction with the model atomic system and proceed to discuss the selection rules governing the dipole transition  matrix element.  We also discuss the experimental results emerging from work by Araoka \textit{et al}.~\cite{Araoka2005a} which conforms with the theoretical prediction for this case. In section \ref{EVCase} we consider the corresponding theory for EVs interacting with the model atomic system and deduce the selection rules governing the dipole transitions in this case as well.  In section \ref{DicEels} we discuss the predictions of our theory in relation to the results of the recent experiment on EELS by Verbeeck \textit{et al}.~\cite{Verbeeck2010} who reported the observation of dichroism in electron energy loss spectroscopy  of thin film magnetised iron samples using EVs.  Section \ref{Concls} contains a summary of the main conclusions regarding the comparison between the two vortex interactions with matter and provides further comments. Some details are consigned to Appendices.

\section{Optical and electron vortex mode functions}\label{OVEVMF}
\subsection{Bessel modes}\label{Besselmodes}
For both types of vortices we shall concentrate on the simplest type of vortex function, namely the Bessel mode.  As a solution to both the Helmholtz and Schr\"{o}dinger equations the Bessel beam is a suitable carrier for orbital angular momentum for both optical and electronic systems; choosing solutions involving only Bessel functions of the first kind gives a mode with zero intensity along the optical axis, as required to support the phase singularity arising from the vortex phase factor $e^{il\phi}$.

In the first case of the optical vortex  the mode function is characterised by a transverse electric field which is a solution of the electromagnetic vector Helmholtz equation.  In cylindrical polar coordinates $\mathbf{r}=(\rho, \phi,z)$ the optical vortex mode function is 
\begin{equation}
\mathbf{E}(\mathbf{r}, t)=E_0 J_l(k_{\perp} \rho) e^{i k_{z} z}e^{i l \phi} e^{-i \omega t}\hat{\pmb{\varepsilon}},
\label{EfieldBessOpt}
\end{equation}
where $E_0$ is the mode amplitude and the unit vector $\hat{\boldsymbol{\varepsilon}}$ denotes the wave polarisation vector.  Since we are mainly concerned with orbital angular momentum, we shall assume that $\hat{\boldsymbol{\varepsilon}}$ stands for linear polarisation, unless stated otherwise. The radial function  $J_l(k_{\perp} \rho)$ is the Bessel function of the first kind of order $l$, where $l$ is the winding number. The wavevectors $k_{\perp}$ and $k_z$ stand for in-plane and axial wavevector variables respectively, such that $k^2=k_{\perp}^2 + k_z^2$, and $\omega$ is the frequency of the light. For the optical case, the transverse vector potential is related to the vortex electric field of \eqaref{EfieldBessOpt} by 
\begin{equation}
\mathbf{A}(\mathbf{r},t) = -\frac{i}{\omega}\mathbf{E}(\mathbf{r},t)
\end{equation}

The electron vortex mode is characterised by a wavefunction $\psi(\mathbf{r},t)$, which is a solution of the scalar Helmholtz equation emerging from the Schr\"{o}dinger equation in cylindrical polar coordinates.  For convenience and for ease of comparison we retain the same symbols, including  wavevector and frequency variables.  We have
\begin{equation}
\psi(\mathbf{r},t)=N J_l(k_{\perp} \rho) e^{i k_{z} z}e^{i l \phi} e^{-i \omega t},
\label{WavefnBessElec}
\end{equation}  
where $N$ is a suitable normalisation constant and $\omega=\mathcal{E}/\hbar$ with $\mathcal{E}$ the mode energy. For both types of vortex mode the vorticity resides in the phase factor $e^{il\phi}$.  Note that the two vortex functions described above are similar in appearance and formally have the same spatial and temporal distributions; however their physical characteristics differ markedly, firstly in the scales of variation and, secondly,  since they describe very different phenomena, their respective interactions with matter differ significantly. For the optical vortex case, the coupling is via the minimal coupling prescription, leading to interaction terms of the form $\mathbf{p}\cdot\mathbf{A}(\mathbf{r},t)$ representing the interaction of the transverse vortex vector potential with the momentum, $\mathbf{p}$, of each of the atomic particles. In contrast, the leading interaction of the charged electron vortex is a coupling to the two-particle atomic system via the Coulomb interaction, with scalar potential $\Phi({\bf r'})$ coupling to each of the atomic constituents. We have
\begin{equation}
\Phi(\mathbf{r'})=-\frac{e}{4\pi\epsilon_0}\frac{1}{|\mathbf{r}_v-\mathbf{r'}|}
\end{equation}
as the Coulomb potential at $\mathbf{r'}$ due to the electron vortex with position variable $\mathbf{r}_v$.

The main task is to explore, by direct analysis, in each case how the transfer of OAM can occur in transitions between states of the vortex beam and the two-particle system, and deduce the selection rules governing those processes. To this end we use a rigorous canonical theory based on the Lagrangian which leads us to the total Hamiltonian of the unperturbed states and their interactions.   

\subsection{Total Lagrangian and Hamiltonian}\label{TotLH}
\begin{figure}%
\includegraphics[width=0.7\columnwidth]{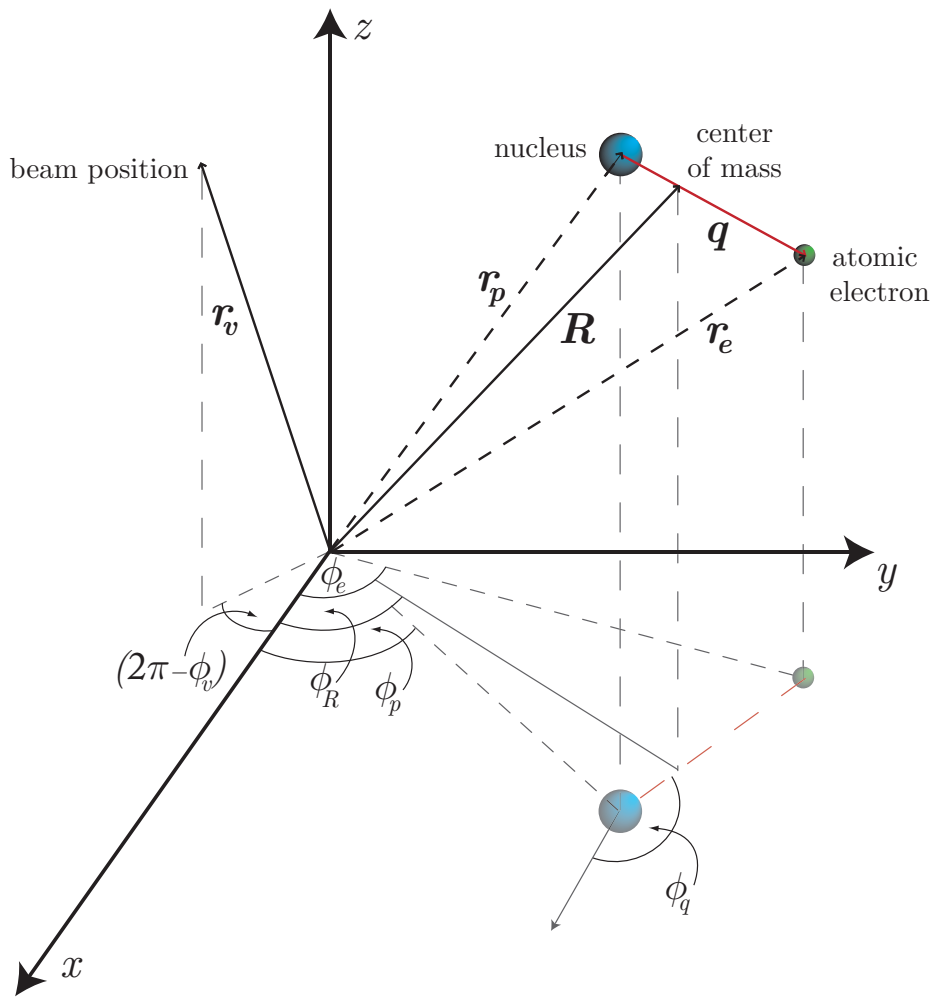}%
\caption{The relevant coordinate frames in the description of the interaction between a two-particle neutral system and Bessel-type optical or electron vortex beam (schematic). 
 The vortex position variable, $\mathbf{r}_v$ relative to the laboratory frame is given in cylindrical coordinates ;  $\mathbf{R}$ is the position variable of the atomic center of mass, and $\mathbf{q}$ stands for the position variable of the internal (electron-type) motion.  The projections of the three position vector variables on the $xy$ plane are seen to have azimuthal anglular positions $\phi_v$, $\phi_R$, and $\phi_q$ respectively.}%
\label{diagram}%
\end{figure}

With the laboratory coordinate system taken to be centered at the vortex origin, as shown in Fig.~\ref{diagram}, we shall assume that the two-particle system is hydrogenic, consisting of an electron of charge $-e$, mass $m_e$, position vector $\mathbf{r}_e$ and momentum $\mathbf{p}_e$; and a nucleus of mass $m_p$, charge $e$, position vector $\mathbf{r}_{p}$ and momentum $\mathbf{p}_{p}$ - all relative to the (laboratory) coordinate system in which the vortices are defined with position variable $\mathbf{r}_v$, and such that all position vectors are in cylindrical coordinates $\mathbf{r}_{\alpha}=(\rho_{\alpha}, \phi_{\alpha},z_{\alpha})$. In this frame of reference the two-particle system possesses a total charge density given by
\begin{equation}
\tilde{\rho}(\mathbf{r})=e\delta(\mathbf{r}-\mathbf{r}_p)-e\delta(\mathbf{r}-\mathbf{r}_e)
\end{equation}
The Lagrangian of the two-particle system in the presence of the interaction with \textit{both} vortex beams can be written as
\begin{equation}
L=L^{0}_{\text{atomic}}+L^{0}_{\text{vortex}}+L^{\text{int}}
\end{equation}
Explicitly we have (following standard Lagrangian techniques as in, for example \cite{CohenTannoudji1989})
\begin{gather}
L^{0}_{\text{atomic}}=\frac{1}{2}m_e\mathbf{\dot{r}}_e^2+\frac{1}{2}m_p\mathbf{\dot{r}}_p^2+\frac{e^2}{4\pi\varepsilon_0}\frac{1}{|\mathbf{r}_p-\mathbf{r}_e|};\\
L^{0}_{\text{vortex}}=\frac{\varepsilon_0}{2}\int\left(\dot{\mathbf{A}}^2(\mathbf{r},t)-c^2\left(\nabla\times\mathbf{A}(\mathbf{r},t)\right)^2 \right)d^3r +\frac{1}{2}m_e\dot{\mathbf{r}}^2_v;\\
L^{\text{int}}=e\mathbf{\dot{r}}_p\cdot\mathbf{A}(\mathbf{r}_p)-e\mathbf{\dot{r}}_e\cdot\mathbf{A}(\mathbf{r}_e)-\int\tilde{\rho}(\mathbf{r'})\Phi(\mathbf{r'})d^3r'.
\end{gather}
Thus $L^{0}_{\text{atomic}}$ is simply the sum of the kinetic energies minus the inter-particle Coulomb potential energy of the atomic particles, and $L^{0}_{\text{vortex}}$ is the zero order energies of the optical and electron vortices.  $L^{\text{int}}$ contains the interaction of the optical vortex vector potential with each of the atomic particles, as well as the interaction of the electron vortex Coulomb potential with the charge density of the atomic particles.

The formalism so far is such that the coordinates of the two-particles forming the atom are expressed entirely relative to the laboratory frame.  However, to be able to discuss  transitions involving the internal states of the atom, we proceed by expressing the atomic motion in terms of the gross (center of mass-type) motion and the internal (electronic-type) motion.  This is achieved by the following transformation 
\begin{equation}
\mathbf{q}=\mathbf{r}_e-\mathbf{r}_p;\;\;\;\;\;\; \mathbf{R}= \frac{m_e\mathbf{r}_e+m_p\mathbf{r}_p}{M},
\label{transfm}
\end{equation}
where $M=m_e+m_p$ is the total mass.  We now have $\mathbf{q}$ as the internal coordinate representing electronic-type motion about the nuclear position $\mathbf{r}_p$, while $\mathbf{R}$ is the coordinate of the center of mass in the laboratory frame. In terms of $\mathbf{q}$ and $\mathbf{R}$ the particle position variables are given by
\begin{equation}
\mathbf{r}_e=\mathbf{R}+\frac{m_p}{M}\mathbf{q};\;\;\;\;\;\; \mathbf{r}_p= \mathbf{R}-\frac{m_e}{M}\mathbf{q}
\label{transfmed}
\end{equation}
Substituting for $\mathbf{r}_e$ and $\mathbf{r}_p$ using \eqaref{transfmed}, the Lagrangian becomes 
\begin{multline}
L=\frac{1}{2}m_e\left(\mathbf{\dot{R}}+\frac{m_p}{M}\mathbf{\dot{q}}\right)^2+\frac{1}{2}m_{p}\left(\mathbf{\dot{R}}-\frac{m_e}{M}\mathbf{\dot{q}}\right)^2-e\left(\mathbf{\dot{R}}+\frac{m_p}{M}\mathbf{\dot{q}}\right)\cdot\mathbf{A}(\mathbf{r}_e) +e\left(\mathbf{\dot{R}}-\frac{m_e}{M}\mathbf{\dot{q}}\right)\cdot\mathbf{A}(\mathbf{r}_p)\\
+\frac{e^2}{4\pi\varepsilon_0}\frac{1}{|\mathbf{q}|}+\frac{e^{2}}{4\pi\epsilon_0}\left(\frac{1}{|\mathbf{r}_v-\mathbf{r}_p|}-\frac{1}{|\mathbf{r}_v-\mathbf{r}_{e}|}\right)+L^{0}_{\text{vortex}}
\end{multline}
where, for convenience at this stage, we have left unchanged the implicit dependence of the vector potential $\mathbf{A}(\mathbf{r})$ on the original particle coordinates, and have evaluated the integral involving $\Phi(\mathbf{r'})$.  We note again that $\mathbf{r}_v$ denotes the vortex position variable.  We now determine the momenta $\mathbf{p}_R$ and $\mathbf{p}_q$ canonically conjugate to $\mathbf{R}$ and $\mathbf{q}$ respectively.  We have after rearrangements 
\begin{gather}
\mathbf{p}_R=\frac{\partial L}{\partial\mathbf{\dot{R}}}= M\mathbf{\dot{R}}-e({\mathbf {\delta A}})\label{cmp}\\
\mathbf{p}_q=\frac{\partial L}{\partial\mathbf{\dot{q}}}=\mu\mathbf{\dot{q}}-e({\mathbf {\Sigma A}})\label{relp},
\end{gather}
where we have introduced the short hand notation
\begin{gather}
(\delta\mathbf{A})=\mathbf{A}(\mathbf{r}_e)-\mathbf{A}(\mathbf{r}_p)\\
(\Sigma\mathbf{A})=\frac{m_p}{M}\mathbf{A}(\mathbf{r}_e)+\frac{m_e}{M}\mathbf{A}(\mathbf{r}_p).
\end{gather}
The momentum canonically conjugate to $\mathbf{r}_v$ is $\mathbf{p}_v=m_e\dot{\mathbf{r}}_v$ and the momentum density canincally conjugate to $\mathbf{A}$ is $\mathbf{\Pi}=\epsilon_0\dot{\mathbf{A}}$.  The Hamiltonian representing the two-particle atom in the presence of the interaction with the two vortices is given by the standard expression
\begin{equation}
H=\mathbf{p}_R\cdot\mathbf{\dot{R}}+\mathbf{p}_q\cdot\mathbf{\dot{q}}+\mathbf{p}_v\cdot\mathbf{\dot{r}}_v+
\int\left(\boldsymbol{\Pi}\cdot\dot{\mathbf{A}} \right)d^3r-L.
\end{equation}
Substituting for all the canonical momenta and momentm density we find that a number of terms cancel to yield the result
\begin{equation}
H=\frac{1}{2}M\mathbf{\dot{R}}^2+\frac{1}{2}\mu\mathbf{\dot{q}}^2 -\frac{e^2}{4\pi\varepsilon_0}\frac{1}{|\mathbf{q}|}-\frac{e^{2}}{4\pi\epsilon_0}\left(\frac{1}{|\mathbf{r}_v-\mathbf{r}_p|}-\frac{1}{|\mathbf{r}_v-\mathbf{r}_{e}|}\right)+H^{0}_{\text{v}},
\label{canonH}
\end{equation}
where $\mu=m_em_p/M$ is the reduced mass and $H^0_{\text{v}}$ is given below in \eqaref{HintV}.  That the Hamiltonian in \eqaref{canonH} incorporates the interactions with the two vortices can be seen on eliminating the particle velocities in favour of the canonical momenta using.  We have
\begin{equation}
H=\frac{[\mathbf{p}_q+e(\Sigma \mathbf{A})]^2}{2\mu}+\frac{[\mathbf{p}_R+e(\delta\mathbf{ A})]^2}{2M}
-\frac{e^2}{4\pi\varepsilon_0}\frac{1}{|\mathbf{q}|}-\frac{e^{2}}{4\pi\epsilon_0}\left(\frac{1}{|\mathbf{r}_v-\mathbf{r}_p|}-\frac{1}{|\mathbf{r}_v-\mathbf{r}_{e}|}\right)+H^{0}_{\text{v}}.
\end{equation}
Expanding the squares in the first two terms,  we find
\begin{equation}
H=\frac{\mathbf{p}_q^2}{2\mu}-\frac{e^2}{4\pi\varepsilon_0}\frac{1}{|\mathbf{q}|}+\frac{\mathbf{p}_R^2}{2M}
+\frac{e}{\mu}\mathbf{p}_q\cdot(\Sigma \mathbf{A})+\frac{e}{M}\mathbf{p}_R\cdot(\delta \mathbf{A})
-\frac{e^{2}}{4\pi\epsilon_0}\left(\frac{1}{|\mathbf{r}_v-\mathbf{r}_p|}-\frac{1}{|\mathbf{r}_v-\mathbf{r}_{e}|}\right)+H^{0}_{\text{v}}
\label{expand}
\end{equation}
where we have retained only the terms linear in $\mathbf{A}(\mathbf{r})$, since for the optical vortex case our main concern is with transitions involving single optical vortex photons. Finally, we concentrate on the electric dipole approximation and expand the functions of $(\mathbf{R}-\frac{m_e}{M}\mathbf{q})$ and $(\mathbf{R}+\frac{m_p}{M}\mathbf{q})$ in \eqaref{expand} about the center of mass coordinate $\mathbf{R}$, to find
\begin{gather}
\frac{1}{|\mathbf{r}_v-\mathbf{r}_p|}-\frac{1}{|\mathbf{r}_v-\mathbf{r}_{e}|}\approx \frac{\mathbf{q}\cdot(\mathbf{r}_v-\mathbf{R})}{|\mathbf{r}_v-\mathbf{R}|^3} + \mathcal{O}(\mathbf{q^2});\\
\mathbf{A}(\mathbf{r}_e)\approx\mathbf{A}(\mathbf{R}) +\frac{m_p}{M}(\mathbf{q}\cdot\nabla)\mathbf{A}(\mathbf{R})+ \mathcal{O}(\mathbf{q^2});\\
\mathbf{A}(\mathbf{r}_p)\approx\mathbf{A}(\mathbf{R}) -\frac{m_e}{M}(\mathbf{q}\cdot\nabla)\mathbf{A}(\mathbf{R})+ \mathcal{O}(\mathbf{q^2}),
\label{dipexp}
\end{gather}
Retaining only those terms up to first order in $\mathbf{q}$, as is consistent with the dipole approximation, we can then write
\begin{equation}
H=H^0_q+H^0_{\text{cm}}+H^0_{\text{v}}+ H^{\text{int(q)}}_{OV}+H^{\text{int(R)}}_{OV}+H^{\text{int}}_{EV},
\end{equation}
where the individual terms are expressed as
\begin{gather}
H^0_q=\frac{\mathbf{p}_q^2}{2\mu}-\frac{e^2}{4\pi\varepsilon_0}\frac{1}{|\mathbf{q}|};\\
H^0_{\text{cm}}=\frac{\mathbf{p}_R^2}{2M};\\
H^0_{\text{v}}=\frac{\varepsilon_0}{2}\int\left(\dot{\mathbf{A}}^2(\mathbf{r})+c^2(\nabla\times\mathbf{A}(\mathbf{r}))^2\right)d^3r+\frac{\mathbf{p}_v^2}{2m_e};\label{HintV}\\
H^{\text{int}(q)}_{\text{OV}}=\frac{e}{\mu}\mathbf{p}_q\cdot\mathbf{A}(\mathbf{R});\label{HintOV1}\\
H^{\text{int}(R)}_{\text{OV}}=\frac{e}{M}\mathbf{p}_R\cdot(\mathbf{q}\cdot\mathbf{\nabla})\mathbf{A}(\mathbf{R});\label{HintOV2}\\
H^{\text{int}}_{\text{EV}}=\frac{e^2}{4\pi\epsilon_0}\frac{\mathbf{q}\cdot(\mathbf{r}_v-\mathbf{R})}{|\mathbf{r}-\mathbf{R}|^{3}}\label{HintEV}
\end{gather}
to dipole order. We can now write the overall Hamiltonian of the system as the sum
\begin{equation}
H=H^0+H^{\text{int}}_{\text{OV}}+H^{\text{int}}_{\text{EV}},
\label{Htotal}
\end{equation}
where $H^0=H^0_q+H^0_{\text{cm}}+H^0_{\text{v}}$ is the zero order Hamiltonian of the overall system , while $H^{\text{int}}_{\text{OV}}$ and $H^{\text{int}}_{\text{EV}}$ are the interaction Hamiltonians of the two-particle system with the optical vortex (including the R\"{o}ntgen interaction given by \eqaref{HintOV2}, c.f.~\cite{VanEnk1994a})  and the electron vortex, respectively.  The total zero order Hamiltonian $H^0$ consists of four separate zero order Hamiltonians representing the distinct sub-systems, namely, $H^0_q$ representing the internal (electronic-type) motion, $H^0_{cm}$ representing the gross (center of mass-type) motion while, as mentioned above, $H^0_{\text{v}}=H^0_{\text{OV}}+H^0_{\text{EV}}$ represents the zero-order Hamiltonian of the free optical vortex and the electron vortex, respectively.  We have
\begin{equation}
H^0=H^0_q+H^0_{\text{cm}}+H^0_{\text{OV}}+H^0_{\text{EV}}.
\label{Hzero}
\end{equation}
In this paper we will be dealing with only one vortex at a time interacting with the model atomic system, so that in applications of \eqaref{Hzero} only three zero-order Hamiltonians are required.  For the optical vortex case $H^0_{\text{EV}}$ is excluded, while in the electron vortex case $H^0_{\text{OV}}$ is excluded.

\subsection{Unperturbed quantum states}\label{UnpertQS}
The selection rules can be deduced from a careful analysis of the transition matrix element evaluated between appropriate final and initial states $\ket{\Psi^{f}}$ and $\ket{\Psi^{i}}$ of the vortex plus the two-particle system.  These are eigenstates of the unperturbed Hamiltonian $H^0$. For the optical case the initial and final states are products of the atomic electron and center of mass eigenstates, and the number state of the optical vortex field:
\begin{equation}
\Ket{\Psi^{i,f}}= \Ket{\vphantom{\Psi^{i,f}}\psi_q^{i,f}(\mathbf{q});\psi_R^{i,f}(\mathbf{R});n^{i,f}},
\label{wavefunctionsop}
\end{equation}
The number states $\ket{n^{i,f}}$ are eigenstates of the optical vortex Hamiltonian for which the vector potential operator is written in terms of annihilation and creation operators, $\hat{a}_{k_{\perp},k_z}$ and $\hat{a}^{\dagger}_{k_{\perp},k_z}$ as follows
\begin{equation}
\mathbf{\hat{A}}(\mathbf{r},t) = \mathbf{A}(\mathbf{r},t)\hat{a}_{k_{\perp},k_z}+\mathbf{A}^{*}(\mathbf{r},t)\hat{a}^{\dagger}_{k_{\perp},k_z}.
\label{QuantMagVecPot}
\end{equation}
For the electron vortex case, the initial and final states are written as product states of the electron vortex wavefunction and the atomic internal electron and center of mass eigenstates, given as
\begin{equation}
\Ket{\Psi^{i,f}}= \Ket{\vphantom{\Psi^{i,f}}\psi_q^{i,f}(\mathbf{q});\psi_R^{i,f}(\mathbf{R});\psi_v^{i,f}(\mathbf{r}_{v})}.
\label{wavefunctionsel}
\end{equation} 

The internal states of the atom $\ket{\psi_q(q,\theta_q,\phi_q)}=\ket{n_q;\ell;m}$ are eigenstates of $\hat{H}^0_q$ and can be formally identified as the well known hydrogenic states, here given in spherical coordinates $(q,\theta_q, \phi_q)$.  Explicitly we have
\begin{equation}
\Ket{\psi_q(\mathbf{q})}=\Ket{\psi_q(q,\theta_q,\phi_q)}=N_{n,\ell,m}Q_{n}(q)P_{{\ell}}^{m}(\cos(\theta_q))e^{im\phi_q}
\label{hydrogenwf}
\end{equation}
where the integer ${\ell}$ is the internal atomic orbital angular momentum (not to be confused with $l$, the vortex OAM quantum number about the beam axis); $m$  is the internal atomic magnetic quantum number (such that $-{\ell}\leq m\leq {\ell}$), and $n$ is the principal quantum number of the internal atomic motion. 

The eigenstates of the center of mass are taken to be product states of both its translational and rotational motion 
\begin{equation}
\Ket{\psi_R(\mathbf{R})}=\Ket{\psi_R(\rho_{R}, \phi_R, z_R)}=\mathcal{R}(\rho_R)e^{iK_{R}\rho_R}e^{i K_zz_R}e^{iL\phi_R}.
\label{comwf}
\end{equation}
where the subscript $R$ indicates center of mass coordinates relative to the laboratory frame.  ${K}_{R}$ and $K_z$ are center of mass wavevectors for the in-plane translational motion and motion along the $z$-axis, such that the total linear momentum of the center of mass is given by $K^2=K_z^2+K_{R}^2$. $L$ is the orbital angular momentum quantum number of the center of mass about the beam axis.  

The vortex states of both the optical vortex and the electron vortex are described above by \eqaref{EfieldBessOpt} and \eqaref{WavefnBessElec} respectively; vectors and their components relating to the vortex beams will henceforth be denoted with the subscript $v$.  In \eqaref{wavefunctionsop} and \eqaref{wavefunctionsel}, and also in what follows, initial and final values of functions or constants are respectively denoted by the superscripts $i$ or $f$, while final values of quantum numbers are indicated by the presence of a dash.

We seek to determine the orbital angular momentum selection rules in processes  involving dipole active transitions due to the interactions between the vortices and the atom such that an exchange of orbital angular momentum occurs between the three subsystems.  This will be done separately for the optical vortex and the electron vortex, in each case  by analysing the complex matrix element $\braket{\Psi^f|\hat{H}^{\text{int}}_{\text{O(E)V}}|\Psi^i}$ and its modulus square; the latter enters the well known formula for Fermi's golden rule leading to the evaluation of the transition rate.  Note, however, that the orbital angular momentum is described by the azimuthal features of the optical vortex field and electron vortex wavefunction.  For this reason only these azimuthal angular components in the matrix elements need be explicitly evaluated.  In this way, the orbital angular momentum selection rules will be made apparent, and the question of transfer of orbital angular momentum between the vortex and the atomic internal dynamics will be answered.

\section{The optical vortex case}\label{OVCase}

The two terms identified as dipole order interaction Hamiltonians, for the OV are given by \eqaref{HintOV1} and \eqaref{HintOV2}.  It can be shown that $H^{\text{int(R)}}_{\text{OV}}$ is in fact the same as the so-called R\"{o}ntgen interaction Hamiltonian \cite{VanEnk1994a}  which couples the center of mass motion and the internal motion via the optical vortex magnetic field $\mathbf{B}$.  We shall not consider the effects of this R\"{o}ntgen interaction Hamiltonian any further here, as it is typically much smaller than the direct coupling with the vector potential , given by $H^{\text{int(q)}}_{\text{OV}}$ \cite{Horsley2005}. We now write $H^{\text{int(q)}}_{\text{OV}}$ in operator form: 
\begin{equation}
\hat{H}^{\text{int}(q)}_{\text{OV}}=-\frac{e}{\mu}\mathbf{p}_q\cdot\mathbf{\hat{A}}(\mathbf{R}).
\end{equation}
The interaction is proportional to the linear momentum operator $\mathbf{p}_{q}$ of the internal atomic motion, and it should be noted that the vector potential operator is evaluated at the center of mass coordinate $\mathbf{R}$, expressed in the laboratory frame of reference (see Fig.~\ref{diagram}). The transition matrix element is given by
\begin{equation}
\mathcal{M}_{\text{OV}}^{fi}=\Braket{\psi^f_q(\mathbf{q});\psi^f_R(\mathbf{R});n^f|\hat{H}^{\text{int}(q)}_{\text{OV}}|\psi^i_q(\mathbf{q});\psi^i_R(\mathbf{R});n^i}.
\label{ovmatrixelement}
\end{equation}
That the interaction Hamiltonian contributes to the electric dipole transition by virtue of the presence of internal momentum operator ${\bf p}_q$, can be seen using the commutator identity
\begin{equation}
\mathbf{p}_q=\frac{i\mu}{\hbar}[\hat{H}^{0}_q,\mathbf{q}].
\end{equation}
In the context of the matrix element we have the standard result
\begin{align}
\Braket{\psi^f_q|\mathbf{p}_q|\psi^i_q}&=\frac{i\mu}{\hbar}\Braket{\psi^f_q|[\hat{H}^{0}_q,\mathbf{q}]|\psi^i_q}\\
&=\frac{i\mu(W_f-W_i)}{\hbar}\Braket{\psi^f_q|\mathbf{q}|\psi^i_q},
\end{align}
where $W_i$ and $W_f$ are hydrogenic-type eigenenergies  of the initial and final hydrogenic-type states  participating in the transition, such that $\hbar^{-1}(W_f-W_i)=\omega$.  It is clear that the interaction in the dipole approximation involves only  the center of mass cylindrical coordinates $(\rho_R, \phi_{R}, z_{R})$.  Substituting this into the transition matrix element we can write
\begin{equation}
\mathcal{M}_{\text{OV}}^{fi}=\frac{i(W_i-W_f)}{\hbar}\Braket{\psi^f_q|\hat{\pmb{\epsilon}}\cdot\mathbf{d}|\psi^i_q}\Braket{\psi^f_R(\mathbf{R});n^f|\hat{A}(\mathbf{R})|\psi^i_R(\mathbf{R});n^i}
\end{equation}
where we have replaced $\mathbf{q}$ by $\mathbf{d}=e\mathbf{q}$, the electric dipole moment vector, and $\hat{A}(\mathbf{R})$ is the scalar operator of the vortex vector potential, the optical polarisation vector  $\hat{\pmb{\epsilon}}$ being incorporated into the dipole matrix element, $\Braket{\hat{\pmb{\epsilon}}\cdot\mathbf{d}}_{fi}$.  Evaluation of the integral of this scalar vortex potential over the initial and final states yields
\begin{multline}
\mathcal{M}_{\text{OV}}^{fi}=\frac{i\mu(W_f-W_i)}{\hbar}\Braket{\psi^f_q|\hat{\pmb{\epsilon}}\cdot\mathbf{d}|\psi^i_q}\Big[\mathcal{A}\delta_{(L, L'-l)}\delta_{(n^i,n^f+1)}\delta(K'_{z}-K_{z}-k_z)\\
-\mathcal{B}\delta_{(L, L'+l)}\delta_{(n^i,n^f-1)}\delta(K'_{z}-K_{z}+k_z)\Big],
\label{cmtransfer1}
\end{multline}
where the factors $\mathcal{A}$ and $\mathcal{B}$ arise from the integration, and are not the same. The conservation of OAM is indicated by the presence of Kronecker deltas $\delta_{(L, L'-l)}$ and $\delta_{(L, L'+l)}$ such that there is an exchange of orbital angular momentum of magnitude $l\hbar$.  The first term indicates the process of absorption of a vortex photon by the center of mass, increasing the orbital angular momentum of the center of mass about the $z$-axis by $l\hbar$, and decreasing the occupancy of the photon field by one.  The second term describes the reverse process, that of emission of a vortex photon by the center of mass. \eqaref{cmtransfer1} embodies conservation of linear as well as angular momentum.  The Dirac delta functions $\delta(K'_{z}-K_{z}-k_z)$ and $\delta(K'_{z}-K_{z}+k_z)$ exhibit the exchange of linear momentum between the the optical vortex and the center of mass such that by absorbing or emitting a vortex photon the linear momentum of the center of mass changes by $\hbar\pm k_z$.

Thus we conclude that it is possible to transfer orbital angular momentum only between the optical vortex and the center of mass,  as previously shown by \cite{VanEnk1994a, Babiker2002}.  The internal electron-type motion is involved in the process of OAM transfer only if the optical vortex beam is circularly polarised, as can be seen by evaluation of the dipole matrix element $\braket{\psi^f_q|\hat{\mathbf{\epsilon}}\cdot\mathbf{d}|\psi^i_q}$, but this has nothing to do with the orbital angular momentum of the light field due to the vortex factor $e^{il\phi_R}$.  In the dipole approximation, it is not possible to transfer orbital angular momentum from the vortex to the internal degrees of freedom of the electron-type motion.  This is in agreement with the results of the experimental investigations by 
 Araoka \textit{et al}.~\cite{Araoka2005a}. In their work, these authors demonstrated that OV light is not specific in its interaction with chiral matter, and that orbital angular momentum may only be exchanged between the optical vortex and the center of mass of the system.  The experimental results also confirm the theoretical predictions of earlier investigations using the PZW Hamiltonian approach \cite{Babiker2002}.  Further experimental results have confirmed that OAM may be transferred between an OV and the rotational motion of an atom \cite{Tabosa1999a}.

\section{The electron vortex case}\label{EVCase}
For the case of the electron vortex interacting with the two-particle system we have the total Hamiltonian
\begin{equation}
\hat{H}=\hat{H}^0+\hat{H}^{\text{int}}_{\text{EV}},
\end{equation}
where here the zero order Hamiltonian contains only those terms relevant to the two-particle system and the electron vortex;
\begin{equation}
\hat{H}^0=\hat{H}^0_q+\hat{H}^0_{\text{cm}}+\hat{H}^0_{\text{EV}}.
\end{equation}
The interaction Hamiltonian $\hat{H}^{\text{Int}}_{EV}$ is given by the Coulomb interaction, expanded to dipole order:
\begin{equation}
\hat{H}^{\text{int}}_{\text{EV}}=\frac{e^2}{4\pi\epsilon_0}\frac{\mathbf{q}\cdot(\mathbf{r}_v-\mathbf{R})}{|\mathbf{r}_v-\mathbf{R}|^{3}},\tag{\ref{HintEV}}
\end{equation}
as derived above.  The relevant matrix element in this case is
\begin{equation}
\mathcal{M}_{\text{EV}}^{fi}=\Braket{\psi_q^{f}(\mathbf{q});\psi_R^{f}(\mathbf{R});\psi_v^{f}(\mathbf{r}_{v})|\hat{H}^{\text{int}}_{EV}|\psi_q^{i}(\mathbf{q});\psi_R^{i}(\mathbf{R});\psi_v^{i}(\mathbf{r}_{v})},
\end{equation}
which can be rewritten as the following product of matrix elements
\begin{equation}
\mathcal{M}_{\text{EV}}^{fi}=\frac{e^2}{4\pi\varepsilon_0}\Braket{\vphantom{\frac{\mathbf{r}_v}{(\mathbf{R})^3}}\psi_q^{f}(\mathbf{q})|\mathbf{q}|\psi_q^{i}(\mathbf{q})}\cdot\Bra{\vphantom{\frac{\mathbf{r}_v}{(\mathbf{R})^3}}\psi_v^{f}(\mathbf{r}_{v});\psi_p^{f}(\mathbf{R})}\frac{\mathbf{r}_{v}-\mathbf{R}}{|\mathbf{r}_{v}-\mathbf{R}|^3}\Ket{\vphantom{\frac{\mathbf{r}_v}{(\mathbf{R})^3}}\psi_v^{i}(\mathbf{r}_{v});\psi_R^{i}(\mathbf{R})}.
\label{EVMatelSplit}
\end{equation}
The first term can be expressed as the dipole matrix element $\Braket{\psi^f_q|\mathbf{d}|\psi^i_q}$, where, as before, the atomic electron dipole moment is given by $\mathbf{d}=e\mathbf{q}$.  The second term in \eqaref{EVMatelSplit}, when evaluated, yields
\begin{equation}
\left<\frac{\mathbf{r}_{v}-\mathbf{R}}{|\mathbf{r}_{v}-\mathbf{R}|^3}\right>_{fi}=\mathcal{C}\left(\frac{\mathbf{\hat{x}}+i\mathbf{\hat{y}}}{2}\right)\delta_{[(L+l), (L'+l'+1)]} +\mathcal{D}\left(\frac{\mathbf{\hat{x}}-i\mathbf{\hat{y}}}{2}\right)\delta_{[(L+l), (L'+l'-1)]} +\mathcal{I}\mathbf{\hat{z}}\delta_{[(L+l), (L'+l')]},
\label{Term2Eval}
\end{equation}
where $\mathcal{C}$, $\mathcal{D}$ and $\mathcal{I}$ are functions which do not contain angular variables and the carets indicate Cartesian unit vectors (See Appendix B for details).  When the dot product in \eqaref{EVMatelSplit} is straightforwardly evaluated, the relevant matrix elements of the dipole moment components between internal states are those of the form $(q_x\pm iq_y)/2$ and $q_z$.  These would result in the familiar selection rules in optical dipole transitions.  The overall matrix element has the form 
\begin{equation}
\mathcal{M}_{\text{EV}}^{fi}=\mathcal{Q}\delta_{[(L+l), (L'+l'+1)]}\delta_{[m, m'-1]}+\mathcal{S}\delta_{[(L+l), (L'+l'-1)]}\delta_{[m, m'+1]}+\mathcal{U}\delta_{[(L+l), (L'+l')]}\delta_{[m, m']},
\label{EVMatel}
\end{equation}
The first term in \eqaref{EVMatel} indicates the possibility of a single unit of orbital angular momentum being transferred from either the center of mass or the electron vortex to the atomic electron, decreasing the total OAM of the combined vortex-center of mass system by one unit, and increasing the magnetic quantum number of the atomic electron by one unit.  The second term describes the reverse process - that of a unit of OAM being transferred from the internal atomic motion to the vortex-center of mass system; the third term indicates the possibility of an interaction in which no OAM is transferred between the three sub-systems.  This is clearly different to the case involving the optical vortex, in which only transfer between the vortex and the center of mass is possible, unless the beam itself is circularly polarised.  This result was previously reported in \cite{Lloyd2012}.

\section{Dichroism in EELS}\label{DicEels}

Predictions of the OAM selection rules for the EV case can be readily put to the test by consideration in relation to a very recent experiment by Verbeeck \textit{et al}.~\cite{Verbeeck2010}, who set out to measure the dichroic signal in the case of L$_{2}$ and L$_{3}$ edges in magnetized iron thin film.  Their experiment essentially involves a single particle transition matrix element, so only the $m$-selection rule is of importance and it is easy to see that the same OAM selection rules would apply if the single particle internal atomic wavefunction in the model system discussed here is replaced by the many-particle electron wavefunctions of the transition metal atom \cite{Cowan1981}.  This is accomplished by replacing $\ket{\psi^i_q}=\ket{n;\ell;m}$ by $\ket{2p^6 3d^n;j;m_j}$, and $\ket{\psi^f_q}=\ket{n;\ell';m'}$ by $\ket{2p^5 3d^{n+1};j';m_j'}$, where $j$ and $m_j$ are the total angular momentum quantum number and associated magnetic quantum number of the many-particle transition metal atom in which the $2p$ core electrons and $3d$ valence electron states are involved in the transition \cite{Thole1992}.

By Fermi's golden rule, the azimuthal angular dependence of the transition rate $\Gamma_{\pm l}$ for the transition process involving by an EV mode having $l=\pm1$, is such that
\begin{equation}
\Gamma_{\pm l}\propto|\mathcal{M}^{l=\pm1}_{\text{EV}}|^2\hat{\rho}_{f},
\label{Fermi}
\end{equation}
where $\hat{\rho}_{f}$ is the density of final states.  As before, we need only consider the angular dependence, since the primary purpose is to examine the role of the OAM of the EV mode in the transition process.  From \eqaref{EVMatel}the internal atomic dipole transitions in question are those for which $m'_j= m_j\pm 1$, for application to the result obtained by Verbeeck \textit{et al}.~\cite{Verbeeck2010}.    These transitions can occur accompanied  by the transfer of one unit of OAM of $l=\pm 1$, either gained or lost by the EV mode.  Since this interaction takes place within iron thin films, a first approximation is to assume that the atomic centers are fixed, ensuring $L=L'=0$, so that all OAM exchange occurs only between the vortex and the internal motion.  The dichroism signal is proportional to the difference between the transition rates for the scenario involving the vortex with OAM equal to $l$ on one hand and that involving the vortex with OAM equal to $-l$ on the other. Relaxing the fixed atom approximation would involve center of mass states in the form of phonons endowed with OAM, which would then participate in the OAM exchange.    

By examining the possible transitions in the L$_{2}$ and L$_{3}$ edges we can determine whether or not we expect to see dichroism based on the transitions the $l=\pm 1$ beams may excite (without, at this point, taking into account density of available final states).   We deal here with many-electron wavefunctions, and the magnetic quantum number affected here is the total angular momentum magnetic quantum number, $m_j$.  The iron L edge corresponds to transitions from the 2p states to 3d states.  This satisfies the selection rule $\Delta \ell=+1$, and interaction with the beam requires $\Delta m_j =\pm 1$.  Since we deal with the dipole approximation, we have the restriction that $\Delta s=0$.  Using a very simple model in which the angular part of the many-electron wavefunction is expressed in terms of spherical harmonics of the form $Y_{j}^{m_j}$ \cite{Condon1951},  we have now a set of 12 allowed transitions - six each allowed for the different beam orbital angular momentum polarisations, with these six further subdivided into two from the L$_2$ edge and four from the L$_3$ edge for both senses of rotation of the beam (see Appendix \ref{mconsid}). Each possible transition in the interaction of the Fe atom with the $l=+1$ beam has a corresponding transition induced by the $l=-1$ beam, and the strengths of these two interactions are the same (full details are given in Appendix \ref{mconsid}).  The total transition rate $\Gamma_{\pm1}$ for the  L edge is given by the sum of the matrix elements over the set of possible intial and final states, from L$_2$ and L$_3$, multiplied by the densities of the final states. Thus, since each $l=+1$ transition  has a corresponding $l=-1$ transition, and the strengths are the same, the square modulus of the matrix elements will be the same for interactions with EVs with $l=\pm1$.  The dichroic signals will arise from the difference in the $m$-level occupancy.  It should be noted that here the $m$-level occupancy does not refer to that of the 3$d$ levels of the transition metals, but the multiplet states of the $2p^53d^{n+1}$.  Because of the strong spin-orbit coupling between the core hole and the rest of the electronic systems, these signals will be sensitive to the spin magnetization as well \cite{Thole1992}.

Verbeeck \textit{et al}.~measured a clear dichroism signal in their experiment \cite{Verbeeck2010}.  Using the analysis above, this can be explained as due to the difference in the density of states such that $\hat{\rho}_{f}\neq \hat{\rho}_{f}$.  Since these processes are dominated by dipole active transitions the dichroism signal can be detected using small angle scattering.  Clearly this is advantageous when compared with the electron energy loss magnetic chiral dichroism technique used by Schattschneider \textit{et al}.~\cite{Schattschneider2008}, which operates using large scattering angles, and where the cross-section for energy loss is much reduced and experiments are more susceptible to noise. 

\section{Comments and Conclusions}\label{Concls}

We have demonstrated by explicit analysis that a transfer of OAM can indeed occur between an EV mode and the internal electronic-type dynamics of matter involving an electric dipole transition.  This contrasts sharply with the case of optical OAM exchange in the interaction of an OV mode with similar systems.  Our predictions that optical vortices cannot exchange OAM with the internal dynamics of an atom in the leading electric dipole interaction are consistent with the experimental finding  of Araoka \textit{et al}.~\cite{Araoka2005a} who demonstrated that OVs are not specific in their interaction with chiral molecules.  This is also consistent with earlier theoretical predictions by Babiker \textit{et al}.~\cite{Babiker2002} who used a complicated PZW technique to analyse the interaction.  This significant finding implies that the EV beam should play a more effective role in magnetic energy loss spectroscopy than that played by ordinary electron beams and, equally significantly,  than the role played by light beams.  It is well recognised that the dipole transition is often the dominant process in most physical systems.  In particular, it should now be possible for EV beams to be used to detect circular dichroic activity in proteins and other biological molecules, allowing useful information to be gained about their secondary structures \cite{Greenfield1969, Chen1972}.  It is expected that the sensitivity and spatial resolution will be high, provided that radiation damage effects can be mitigated.

In view of the similarities of EVs and OVs it is clearly natural to contemplate whether an experiment similar to that by  Araoka \textit{et al}.~\cite{Araoka2005a} on the handedness of processes involving the possibility of OAM exchange with internal dynamics should be carried out for the EV case. As far as the authors are aware, the only experiment to date involving the transfer of OAM of electron vortices is that by Verbeeck \textit{et al}.~\cite{Verbeeck2010}  who specifically dealt with the case $l=\pm 1$ to investigate the electron energy loss spectroscopy signals from magnetised Fe films using EV beams.  The transitions involved are those by core electrons participating in electric dipole allowed transitions between discrete atomic energy levels within a rigid condensed matter background of essentially fixed Fe atoms.  The expectation was that the two EELS signals, one from $l=1$ and the other from $l=-1$, would be different, suggesting that the EELS revealed an intrinsic chirality of the medium.  This was indeed the case in the experiment by Verbeeck \textit{et al}.  The question, however, arises as to the theoretical basis for the observed dichroism.  
  
Our task in this paper has been two-fold. Firstly, we set out to construct the basic theory underlying the coupling of an OV and an EV to matter represented by a two-particle atom as a neutral bound system of two charged particles. The basic theory has been carried out in a manner which incorporates both types of vortex and as far as the atom is concerned, it was necessary to distinguish internal (electronic-type) and gross (center of mass-type) degrees of freedom of the atomic system.  Secondly we proceeded to consider each vortex separately in its interaction with the atomic system. We have shown that, contrary to the case of OVs,  the theory for EVs allows the transfer of OAM to the internal dynamics for electric dipole transitions, as in the experiment by Verbeeck \textit{et al}.~It turns out that this crucial difference between EVs and OVs is attributed to the distinct interaction mechanisms of the vortex with matter.

Our results suggest that the matrix elements involving the orbital angular momentum transfer has the general vector product form of $\braket{e\mathbf{q}}_{fi}\cdot\braket{\mathbf{f}_v(\mathbf{r})}_{fi}$ where $\braket{\mathbf{eq}}_{fi}$ is the dipole transition matrix element between internal atomic states and $\mathbf{f}_v(\mathbf{r})$ is the effective field seen by both the internal motion and center of mass of the atomic system.  It turns out that the condition for the transfer of orbital angular momentum between the vortex beam  and the internal motion of the atom is for $\mathbf{f}_v(\mathbf{r})$ to exhibit  chirality. In the case of EVs, the longitudinal Coulomb interaction couples the dipole moment of the internal motion of the atom to the electric field of the vortex beam, in a similar manner to that in which atomic electrons couple to the optical field of circularly polarised light. In the case of OVs, in which the vortex is characterised by a linearly polarised  transverse vector potential carrying orbital angular momentum, the transition matrix element only depends on the value of the transverse vector potential at the atomic site, hence the dot product is not chiral-specific and no OAM transfer to the internal dynamics is allowed to accompany an electric dipole transition. A  transfer of OAM to both parts of the atomic system in higher multipolar transitions than the dipolar is, however possible for both types of vortex.

Additionally, we note that the transfer of OAM from the EV to the atom may proceed independent of the dynamics of the atom, due to the long range Coulomb interaction involved.  This is in contrast to the case involving OV beams that requires the center of mass of the atom to be a dynamical variable, free to rotate about the beam axis \cite{VanEnk1994a}.

We have analysed the theory for the general case of topological charge $l$. However for the specific case of the Verbeeck et al. experiment, we have demonstrated that although OAM exchange can occur between the EV and matter in electric dipole transitions for the opposite helicities $l = \pm1$, there is no intrinsic difference in the EELS absorption between the two helicities. We have shown that the experiment by Verbeeck \textit{et al}.~displays a dichroism effect via a similar mechanism to the analogous XMCD experiment \cite{Thole1992}.

\section{Acknowledgements}
The authors are grateful to the University of York for the award of a studentship to S.~Lloyd.

\appendix

\section{EV matrix element}
Starting from
\begin{equation}
\mathcal{M}_{\text{eV}}^{fi}=\frac{e^2}{4\pi\varepsilon_0}\Braket{\vphantom{\frac{\mathbf{R}}{(\mathbf{R})^3}}\psi_q^{f}(\mathbf{r}_{q})|\mathbf{q}|\psi_q^{i}(\mathbf{r}_{q})}\cdot\Bra{\vphantom{\frac{\mathbf{R}}{(\mathbf{R})^3}}\psi_v^{f}(\mathbf{r}_{v});\psi_R^{f}(\mathbf{R})}\frac{\mathbf{r}_{v}-\mathbf{R}}{|\mathbf{r}_{v}-\mathbf{R}|^3}\Ket{\vphantom{\frac{\mathbf{R}}{(\mathbf{R})^3}}\psi_v^{i}(\mathbf{r}_{v});\psi_R^{i}(\mathbf{R})}
\tag{\ref{EVMatelSplit}}
\end{equation}
we seek to fully evaluate this matrix element, by considering it as the product of two matrix elements - the first being the dipole matrix element between hydrogenic wavefunction, and the second that of the Coulomb potential between the center of mass states and vortex wavefunctions.  The dipole matrix element is well known, so we will first concentrate on this second, Coulombic matrix element.  This is best evaluated using Cartesian coordinates, so as to compare the result with the known dipole result, and compute their scalar product. Thus,
\begin{equation}
\frac{(\mathbf{r}_{v}-\mathbf{R})}{|\mathbf{r}_{v}-\mathbf{R}|^3} = \frac{(\rho_v\cos(\phi_v)-\rho_R\cos(\phi_R))\mathbf{\hat{x}}+(\rho_v\sin(\phi_v)-\rho_R\sin(\phi_R))\mathbf{\hat{y}}+(z_v-z_R)\mathbf{\hat{z}}}{\left(\mathcal{F}(\rho_v,z_v,\rho_R,z_R)-\mathcal{G}(\rho_v, \rho_R)\cos(\phi_v-\phi_R)\right)^{\frac{3}{2}}}.
\label{vecexp}
\end{equation}
Evaluating this as the matrix element between $\ket{\psi_v^{f,i}(\mathbf{r}_{v});\psi_R^{f,i}(\mathbf{R})}$ requires the use of the substitution $y=(\phi_v-\phi_R)$, in order to express the intergral in \eqaref{EVMatelSplit} in terms of generic integrals of the form
\begin{equation}
\mathcal{Y}_{\alpha}=\int_{0}^{2\pi}\frac{e^{i(l-l'+\alpha)y}}{\left(\mathcal{F}-\mathcal{G}\cos(y)\right)^{\frac{3}{2}}}dy,
\label{Ysubs}
\end{equation}
where $\alpha$ is an integer taking the values $0$, $\pm1$.  The result, after making this substitution, may be written as
\begin{equation}
\left<\frac{\mathbf{r}_{v}-\mathbf{R}}{|\mathbf{r}_{v}-\mathbf{R}|^3}\right>_{fi}=\mathcal{C}\left(\frac{\mathbf{\hat{x}}+i\mathbf{\hat{y}}}{2}\right)\delta_{[(L+l), (L'+l'+1)]} +\mathcal{D}\left(\frac{\mathbf{\hat{x}}-i\mathbf{\hat{y}}}{2}\right)\delta_{[(L+l), (L'+l'-1)]} +\mathcal{I}\mathbf{\hat{z}}\delta_{[(L+l), (L'+l')]}
\tag{\ref{Term2Eval}}
\end{equation}
where 
\begin{gather}
\mathcal{C}=\kappa\mathcal{Y}_{-1}-\lambda\mathcal{Y}_{0},\\
\mathcal{D}=\kappa\mathcal{Y}_{+1}-\lambda\mathcal{Y}_{0},\\
\mathcal{I}=\eta\mathcal{Y}_0,
\end{gather}
with $\kappa$, $\lambda$ and $\eta$ being factors arising from the integration over the remaining (non-azimuthal) degrees of freedom of $\psi_v(\mathbf{r}_{v})$ and $\psi_R(\mathbf{R})$. 

Combining the two parts involved in the dot product of the overall transition matrix elelment we have
\begin{multline}
\left<\frac{\mathbf{r}_{v}-\mathbf{R}}{|\mathbf{r}_{v}-\mathbf{R}|^3}\right>_{fi}\cdot
\Braket{\mathbf{q}}_{fi} = \frac{\mathcal{C}}{2}\Braket{q_x+iq_y}_{fi}\delta_{[(L+l), (L'+l'+1)]}\\+\frac{\mathcal{D}}{2}\Braket{q_x-iq_y}_{fi}\delta_{[(L+l), (L'+l'-1)]}+\mathcal{I}\Braket{q_z}_{fi}\delta_{[(L+l), (L'+l')]}
\label{dipel}
\end{multline}
The evaluation of each of the dipole component matrix elements $\frac{1}{2}\Braket{q_x\pm iq_y}_{if}$ and $\Braket{q_z}_{if}$ is standard and leads to the usual $m$-selection rules.  The overall matrix element can be written as
\begin{equation}
\mathcal{M}_{\text{EV}}^{fi}=\mathcal{Q}\delta_{[(L+l), (L'+l'+1)]}\delta_{[m, m'-1]}+\mathcal{S}\delta_{[(L+l), (L'+l'-1)]}\delta_{[m, m'+1]}+\mathcal{U}\delta_{[(L+l), (L'+l')]}\delta_{[m, m']},
\tag{\ref{EVMatel}}
\end{equation}
where $\mathcal{Q}$,  $\mathcal{S}$ and $\mathcal{U}$ are factors containing the normalisation constants and integrals not involving azimuthal angles.

\section{Application of the selection rules}\label{mconsid}
So as to compare our results with the experimental results of \cite{Verbeeck2010}, we now apply the selection rules derived for the electron vortex interaction with the two-particle system, by considering transitions between certain initial and final states.  The transitions involved are those of core electrons participating in electric dipole allowed transitions between discrete atomic energy levels within a rigid condensed matter background of essentially fixed Fe atoms at equilibrium.  If the center of mass of the atoms in the lattice is allowed to participate in the angular momentum transfer, the motion would involve phonon states endowed with OAM. The electron vortex interaction matrix element is
\begin{equation}
\mathcal{M}_{\text{EV}}^{fi}=\mathcal{Q}\delta_{[(L+l), (L'+l'+1)]}\delta_{[m, m'-1]}+\mathcal{S}\delta_{[(L+l), (L'+l'-1)]}\delta_{[m, m'+1]}+\mathcal{U}\delta_{[(L+l), (L'+l')]}\delta_{[m, m']},
\tag{\ref{EVMatel}}
\end{equation}
The factors $\mathcal{Q}$ and $\mathcal{S}$ can be written as follows
\begin{gather*}
\mathcal{Q}=\mathcal{K}'\mathcal{C}=\mathcal{K}'\left(\kappa\mathcal{Y}_{-1}-\lambda\mathcal{Y}_{0}\right)\\
\mathcal{S}=\mathcal{K}\mathcal{D}=\mathcal{K}\left(\kappa\mathcal{Y}_{+1}-\lambda\mathcal{Y}_{0}\right).
\end{gather*}
where $\mathcal{K}$ and $\mathcal{K}'$ arise in the evaluation of the dipole matrix elements $\frac{1}{2}\Braket{q_x-iq_y}_{if}$.  From the definition of $\mathcal{Y}_{\alpha}$ we can compare these intensity factors for a specific transition.  Choosing $l=\pm1$, and $l'=0$ we can write
\begin{gather}
\mathcal{C}=\left(\kappa\int_{0}^{2\pi}\frac{e^{0}}{\left(\mathcal{F}-\mathcal{G}\cos(y)\right)^{\frac{3}{2}}}dy-\lambda\int_{0}^{2\pi}\frac{e^{i(y)}}{\left(\mathcal{F}-\mathcal{G}\cos(y)\right)^{\frac{3}{2}}}dy\right),\\
\mathcal{D}=\left(\kappa\int_{0}^{2\pi}\frac{e^{0}}{\left(\mathcal{F}-\mathcal{G}\cos(y)\right)^{\frac{3}{2}}}dy-\lambda\int_{0}^{2\pi}\frac{e^{-i(y)}}{\left(\mathcal{F}-\mathcal{G}\cos(y)\right)^{\frac{3}{2}}}dy\right),
\end{gather}
From which it can be seen that $\mathcal{C}=\mathcal{D}^*$.  This choice of $l$ and $l'$ suggests $m'=m+1$ for $l=1$ and $m'=m-1$ for $l=-1$, from some initial magnetic state $m$. This information can be used to find the factors $\mathcal{K}$, $\mathcal{K}'$ for each case.  This now needs generalising to the many-electron wavefunction of the model iron atom.  In the LS coupling regime, the angular part of the wavefunction is given by the product of the spherical harmonics of the occupied states.  The total orbital angular momentum of the atom is $J=L+S$, and associated magnetic quantum number, $m_j$, is the quantity that will be affected by the absorption or emission of a unit of OAM from the electron vortex.  We can express the angular part of the many-electron wavefunction in terms of spherical harmonics of the form $Y_{j}^{m_j}$ (neglecting numerical and phase factors that arise from the coupling of the constituent electrons) \cite{Condon1951}.  The spherical harmonics are normalised such that
\begin{equation}
(-1)^{\ell-m}Y_{\ell}^{-m}=Y_{\ell}^{m*},
\label{sphchrmncs}
\end{equation}
and this also applies to our many-electron spherical harmonics $Y_{j}^{m_j}$. For our purposes, we seek to explain the dichroism  of the L-edge transitions of magnetised iron thin films observed in EELS using electron vortex beams, by Verbeeck et al.~\cite{Verbeeck2010}.  To do this, we first look at the possible transitions that can be excited in the model iron atom by the two beams of opposite orbital angular momenta, $l=\pm1$.  These are summarised in Table \ref{alltrsnt}.  As can be seen, each transition in the $l=+1$ case has a corresponding, similar, transition in the $l=-1$ case, having $m_j^{(-1)}=-m_j^{(+1)}$ and $m'^{(-1)}_j=-m'^{(+1)}_j$.  It is clear that, due to \eqaref{sphchrmncs}, the dipole matrix elements of the corresponding transitions will have the same magnitude (due to the symmetry properties of the Wigner 3-j symbols used to calculate the coupled spherical harmonics, the strengths for the forward and reverse transitions for $l=\pm1$ remain the same \cite{Cowan1981}, so the numerical factors can be safely neglected, as stated above).
\begin{table}%
\begin{tabular}{lllll}
\hline
\hline
$l=+1$        &                                                 & & $l=-1$   &                                                \\
\hline
L$_{2}\,$     &  2p$_{1/2}(m_j=-1/2) \to$ 3d$_{3/2}(m_j=+1/2)$ & $\quad$ & L$_{2}$  &  2p$_{1/2}(m_j=+1/2) \to$ 3d$_{3/2} (m_j=-1/2)$\\
       	      &  2p$_{1/2}(m_j=+1/2) \to$ 3d$_{3/2}(m_j=+3/2)$ & $\quad$ &          &  2p$_{1/2}(m_j=-1/2) \to$ 3d$_{3/2} (m_j=-3/2)$\\
L$_{3}\,$     &  2p$_{3/2}(m_j=-3/2) \to$ 3d$_{5/2}(m_j=-1/2)$ & $\quad$ & L$_{3}$  &  2p$_{3/2}(m_j=+3/2) \to$ 3d$_{5/2} (m_j=+1/2)$\\
        	    &  2p$_{3/2}(m_j=-1/2) \to$ 3d$_{5/2}(m_j=+1/2)$ & $\quad$ &          &  2p$_{3/2}(m_j=+1/2) \to$ 3d$_{5/2} (m_j=-1/2)$\\
        	    &  2p$_{3/2}(m_j=+1/2) \to$ 3d$_{5/2}(m_j=+3/2)$ & $\quad$ &          &  2p$_{3/2}(m_j=-1/2) \to$ 3d$_{5/2} (m_j=-3/2)$\\     
        	    &  2p$_{3/2}(m_j=+3/2) \to$ 3d$_{5/2}(m_j=+5/2)$ & $\quad$ &          &  2p$_{3/2}(m_j=-3/2) \to$ 3d$_{5/2} (m_j=-5/2)$\\
\hline
\hline
\end{tabular}
\caption{Allowed transitions in the L$_2$ and L$_3$ edges for interaction with an electron vortex beam with $l=+1$ and $l=-1$.}
\label{alltrsnt}
\end{table}

The total transition rate (as observed in \cite{Verbeeck2010}) of the L$_2$ and L$_3$ edges will be given by the sum of the transition rates of the individual transitions in each edge.  So, for the L$_2$ edge we have
\begin{multline}
\Gamma_{\text{L}_2}^{l=+1}=\frac{2\pi}{\hbar}|\mathcal{C}|^2\bigg(|\Braket{2p_{1/2},m_j=-1/2|\mathbf{q}|3d_{3/2},m_j=+1/2}|^2 \hat{\rho}_{3d_{3/2}(m_j=+1/2)}\\
 + |\Braket{2p_{1/2}(m_j=+1/2)|\mathbf{q}|3d_{3/2},m_j=+3/2}|^2 \hat{\rho}_{3d_{3/2}(m_j=+3/2)}\bigg),
\end{multline}
and
\begin{multline}
\Gamma_{\text{L}_2}^{l=-1}=\frac{2\pi}{\hbar}|\mathcal{D}|^2\bigg(|\Braket{2p_{1/2},m_j=+1/2|\mathbf{q}|3d_{3/2},m_j=-1/2}|^2 \hat{\rho}_{3d_{3/2}(m_j=-1/2)}\\
 + |\Braket{2p_{1/2},m_j=-1/2|\mathbf{q}|3d_{3/2},m_j=-3/2}|^2 \hat{\rho}_{3d_{3/2}(m_j=-3/2)}\bigg),
\end{multline}
which will be equal as long as $\hat{\rho}_{3d_{3/2}(m_j=+1/2)}=\hat{\rho}_{3d_{3/2}(m_j=-1/2)}$ and $\hat{\rho}_{3d_{3/2},m_j=+3/2}=\hat{\rho}_{3d_{3/2},m_j=-3/2}$, as we have already established that $|\mathcal{C}|^2=|\mathcal{D}|^2$.  The same argument applies to the L$_3$ edge.  Thus we conclude that the observed dichroism is due solely to the distribution of electrons in the magnetised iron, and not due to the mechanism of the interaction with the electron vortex.

\bibliography{References}{}

\end{document}